\documentstyle[prd,aps,epsfig]{revtex}

\tighten  
\begin{document}
\draft

\def\simgt{\mathrel{\raise.3ex\hbox{$>$\kern-.75em\lower1ex\hbox{$\sim$}}}}
\def\simlt{\mathrel{\raise.3ex\hbox{$<$\kern-.75em\lower1ex\hbox{$\sim$}}}}
\newcommand{\s}{\mbox{$\sigma$}}
\newcommand{\bi}[1]{\bibitem{#1}}
\newcommand{\fr}[2]{\frac{#1}{#2}}
\newcommand{\gm}{\mbox{$\gamma_{\mu}$}}
\newcommand{\gn}{\mbox{$\gamma_{\nu}$}}
\newcommand{\Le}{\mbox{$\fr{1+\gamma_5}{2}$}}
\newcommand{\R}{\mbox{$\fr{1-\gamma_5}{2}$}}
\newcommand{\GD}{\mbox{$\tilde{G}$}}
\newcommand{\gf}{\mbox{$\gamma_{5}$}}
\newcommand{\Ima}{\mbox{Im}}
\newcommand{\Rea}{\mbox{Re}}
\newcommand{\Tr}{\mbox{Tr}}
\newcommand{\psl}{\slash{\!\!\!p}}
\newcommand{\cp}{\;\;\slash{\!\!\!\!\!\!\rm CP}}
\newcommand{\qq}{\langle \ov{q}q\rangle}

\newcommand{\nc}{\newcommand}
\nc{\al}{\alpha}
\nc{\ga}{\gamma}
\nc{\de}{\delta}
\nc{\ep}{\epsilon}
\nc{\ze}{\zeta}
\nc{\et}{\eta}
\renewcommand{\th}{\theta}
\nc{\Th}{\Theta}
\nc{\ka}{\kappa}
\nc{\la}{\lambda}
\nc{\rh}{\rho}
\nc{\si}{\sigma}
\nc{\ta}{\tau}
\nc{\up}{\upsilon}
\nc{\ph}{\phi}
\nc{\ch}{\chi}
\nc{\ps}{\psi}
\nc{\om}{\omega}
\nc{\Ga}{\Gamma}
\nc{\De}{\Delta}
\nc{\La}{\Lambda}
\nc{\Si}{\Sigma}
\nc{\Up}{\Upsilon}
\nc{\Ph}{\Phi}
\nc{\Ps}{\Psi}
\nc{\Om}{\Omega}
\nc{\ptl}{\partial}
\nc{\del}{\nabla}
\nc{\ov}{\overline}

\nc{\be}{\begin{equation}}
\nc{\ee}{\end{equation}}
\nc{\bea}{\begin{eqnarray}}
\nc{\eea}{\end{eqnarray}}

\nc{\newcaption}[1]{\centerline{\parbox{5.6in}{\caption{#1}}}}

%\begin{flushright}
%TPI-MINN--02/38\\
%UMN-TH-2109/02\\
%SUSX-TH/02-019\\
%UVic-th/2002-01\\
%DAMTP-2002-109\\
%hep-ph/0208257\\
%\end{flushright}

\twocolumn[\hsize\textwidth\columnwidth\hsize\csname 
@twocolumnfalse\endcsname  

\title{Hadronic EDMs, the Weinberg Operator, and Light Gluinos}

\author{Durmu{\c s} Demir$^{1}$,
 Maxim Pospelov$^{2,3}$ and  Adam Ritz$^{4}$}

\address{$^{1}$Theoretical Physics Institute, University of Minnesota,\\
Minneapolis, MN 55455, USA} 
\address{$^{2}$Centre for Theoretical Physics, CPES, 
               University of Sussex, Brighton BN1~9QJ, UK} 
\address{$^{3}$ Department of Physics and Astronomy, 
University of Victoria,
     Victoria, BC, V8P 1A1 Canada } 
\address{$^{4}$Department of Applied Mathematics and Theoretical
         Physics, Centre for Mathematical Sciences,\\ University of
         Cambridge, Wilberforce Rd., Cambridge CB3 0WA, UK}

\date{\today}

\maketitle

\begin{abstract}
We re-examine questions concerning the contribution of the three-gluon 
Weinberg operator to the electric dipole moment of the neutron, 
and provide several QCD sum rule--based arguments that the 
result is smaller than -- but nevertheless consistent with --
estimates which invoke naive 
dimensional analysis. We also point out a regime of the MSSM parameter 
space with light gluinos for which this operator provides the
dominant contribution to the neutron electric dipole moment due to 
enhancement via the dimension five color electric dipole moment of the gluino.
\end{abstract}

\vskip2pc]

\section{Introduction and Summary}

New sources of $CP$ violation in supersymmetric extensions of the
standard model are highly constrained by the null experimental 
results for the electric dipole moments (EDMs) of neutrons and 
heavy atoms \cite{exp,K}. Typically, when the superpartners
have an electroweak scale mass, $\Lambda_W$, the additional $CP$
violating phases are constrained to be of $O(10^{-2})$. When  
confronted with the natural expectation that the size of these 
phases in the soft-breaking sector should be of order one, this creates a 
problem for weak-scale supersymmetry (SUSY). 

The interactions which generate EDMs  are described by a
$CP$-odd effective Lagrangian, induced at 1GeV by integrating
out heavy standard model particles and superpartners, which
contains a series of operators of increasing dimension. The leading 
$\th$--term,
\be
{\cal L}^{[4]}_{\rm eff}= \frac{g_s^2}{32\pi^{2}}\ \bar\theta\  G^{a}_{\mu\nu}
\widetilde{G}^{a}_{\mu\nu}
\ee
has dimension four, and an arbitrary value for $\bar\theta$ 
constitutes the usual strong $CP$ problem as its contribution to EDMs is
unsuppressed by any heavy scale. Moreover, the  existence of additional 
$CP$-odd phases in the soft-breaking sector of the 
MSSM aggravates this problem by inducing a large
additive renormalization of $\bar \theta$ that survives in the
decoupling limit. The conventional `cure' -- the Peccei-Quinn
mechanism %\cite{PQ} 
-- eliminates $\bar\theta$ and leaves the dimension five
quark EDMs and color EDMs (CEDMs),
\bea
 {\cal L}^{[5]}_{\rm eff} &=& -\frac{i}{2} 
\sum_{i=e,u,d,s} d_i\ \overline{\psi}_i (F\sigma)\gamma_5
\psi_i  \nonumber\\
 && \;\;\;\; - \frac{i}{2} \sum_{i=u,d,s} \widetilde{d}_i\ \overline{\psi}_i
(G\sigma)\gamma_5 \psi_i,
\eea
and the Weinberg operator \cite{wein1},
\be
 {\cal L}^{[6]}_{\rm eff}= \frac{1}{3} w\  f^{a b c} G^{a}_{\mu\nu}
\widetilde{G}_{\nu \beta}^{b}  G^{c}_{\beta\mu}, \label{wo}
\ee
as the dominant mediators of $CP$ violation from  the soft
breaking sector to the observables. Note that although the quark (C)EDMs
have dimension five, chiral symmetry requires that the corresponding
coefficients are proportional to a light quark mass, and 
thus $d_i$, $\widetilde{d}_i$, and $w$ generically scale 
in the same way with the overall SUSY breaking scale.

Extracting constraints on the underlying $CP$-odd phases thus requires
quantitative knowledge of the dependence of observable EDMs on 
$d_i$, $\widetilde{d}_i$, and $w$ normalized at the hadronic scale. 
Recently, the dependence on $d_i$ and $\widetilde{d}_i$ has been
determined more precisely using QCD sum rules \cite{PR}, and in this
note we turn our attention to the Weinberg operator. Although
rather intractable within the standard framework, we will present 
several sum--rule based estimates. The resulting
preferred range for the neutron EDM,
\be
  d_n(w) = e\,(10 - 30)~ {\rm MeV}\, w(1\,{\rm GeV})\,, \label{res}
\ee
is a factor of two smaller than conventional estimates 
\cite{wein1,wein2} using `naive dimensional analysis' (NDA) \cite{nda}. This
moderate suppression can be understood through the appearance of combinatoric
factors which are not accounted for within NDA. However, while our result 
for $d_n(w)$ is smaller than the NDA estimate, and 
thus $d_n(d_i,\widetilde{d}_i)$ generally 
dominates the contributions to $d_n$, there is a regime in which
$d_n(w)$ is important as it is generated rather differently 
from the quark (C)EDMs within the MSSM.

In order to explain this point recall, first of all, that there 
are several generic `strategies' for 
curing the SUSY $CP$ problem. The first is to require that the 
superpartners are heavy enough to suppress all operators of dim$\ge5$ 
generated at the SUSY threshold. This decoupling is usually
applied to sfermions of the first two generations only, in order to avoid
problems with fine-tuning in the Higgs sector. However, this
approach is only partially successful as relatively large EDMs
may be generated through higher loops \cite{twoloop} or through 
four-fermion operators induced by Higgs exchange \cite{LP}.
Secondly, one could conceive of a universal conspiracy leading 
to cancellations between different contributions \cite{cancel}, but
this is difficult to reconcile with the null results for all three
types of EDM measurement (neutron, paramagnetic and
diamagnetic atoms) that a priori have different phase dependence \cite{FOPR}.
A third, perhaps more elegant, option is to invoke  an 
exact $CP$ or parity at some high-energy scale and specify the mechanisms that 
break supersymmetry in such a way that all the relevant soft breaking
parameters are rendered real. This could also be one way of obviating the
need for axion relaxation \cite{CP-P}. However, some of these scenarios 
may face problems when confronted with the large $CP$ violation that is by now 
well-documented in the $B$-meson system \cite{Bcp}. 

Given these difficulties, one may pursue another option which is
to suppress the SUSY contributions by creating some (mild) 
hierarchies between the soft breaking parameters in order to suppress 
the EDMs generated at one-loop. Notably, in the limit where 
gauginos are much lighter than the sfermions, all one-loop
contributions to the EDMs of light quarks and
the electron take the following form: 
\be 
 d_i({\rm one~loop}) \sim ({\rm loop ~ factor}) 
 \times \frac {m_i}{m_{{\rm sf}}^4}{\rm Im}(m_\lambda A),
\label{d_i}
\ee 
with a similar expression for $d_i$ induced by the relative phase
of $\mu$ and $m_\lambda$.  Here $i=e,u,d,s$, and $m_{{\rm sf}}$ stands
for a generic sfermion mass. It is easy to see that 
as $m_\lambda \rightarrow 0$ the expression
(\ref{d_i}) for $d_i$ vanishes.  Thus a mild hierarchy $m_\lambda \sim
(10^{-3} - 10^{-2})\,m_{{\rm sf}}$ would appear to 
be sufficient to evade the SUSY $CP$
problem \cite{Farrar,Austrians}. In slightly different language, it 
follows from (\ref{d_i}) that in this regime the quark EDMs are demoted
to dimension seven operators and thus are relatively harmless.

While the quark EDMs are suppressed by this hierarchy, we emphasize
that sending $m_\lambda$ down to hadronic scales actually {\it
enhances} the neutron EDM via the generation of the Weinberg operator.
The main point is that the gluino CEDM,
\be
{\cal{L}}_{\lambda}= \frac{1}{4} \widetilde{d}_{\lambda}\   f^{abc}
\overline{\lambda^b} \sigma\cdot G^a \gamma_5 \lambda^c,
 \label{gcedm}
\ee
can be induced by a top-stop loop  \cite{apostolos}, leading to
\be
\widetilde{d}_{\lambda}({\rm one~loop}) \sim ({\rm  loop ~ factor})
\times\fr{m_t^2}{m_{\rm sf}^4}{\rm Im}(A_t-\mu^*\cot \beta ) 
\ee 
in a basis in which the gluino mass is real. Thus $\widetilde{d}_{\lambda}$
is a genuine dimension five operator, $\widetilde{d}_{\lambda} \sim
1/\Lambda_W$, for $A_t\sim \mu\sim m_{\rm sf}\sim \Lambda_W$. It follows that
for $m_{\lambda} \sim \Lambda_{\rm hadr}$, the gluino takes part
in the strong interactions and contributes to the energy density of
all hadrons. Consequently the neutron EDM is unsuppressed by any
additional scale, and at a crude level $d_n  \sim e
\widetilde{d}_{\lambda} \sim 1/\Lambda_W $.

\begin{figure}[t]
\begin{center}
\epsfig{file=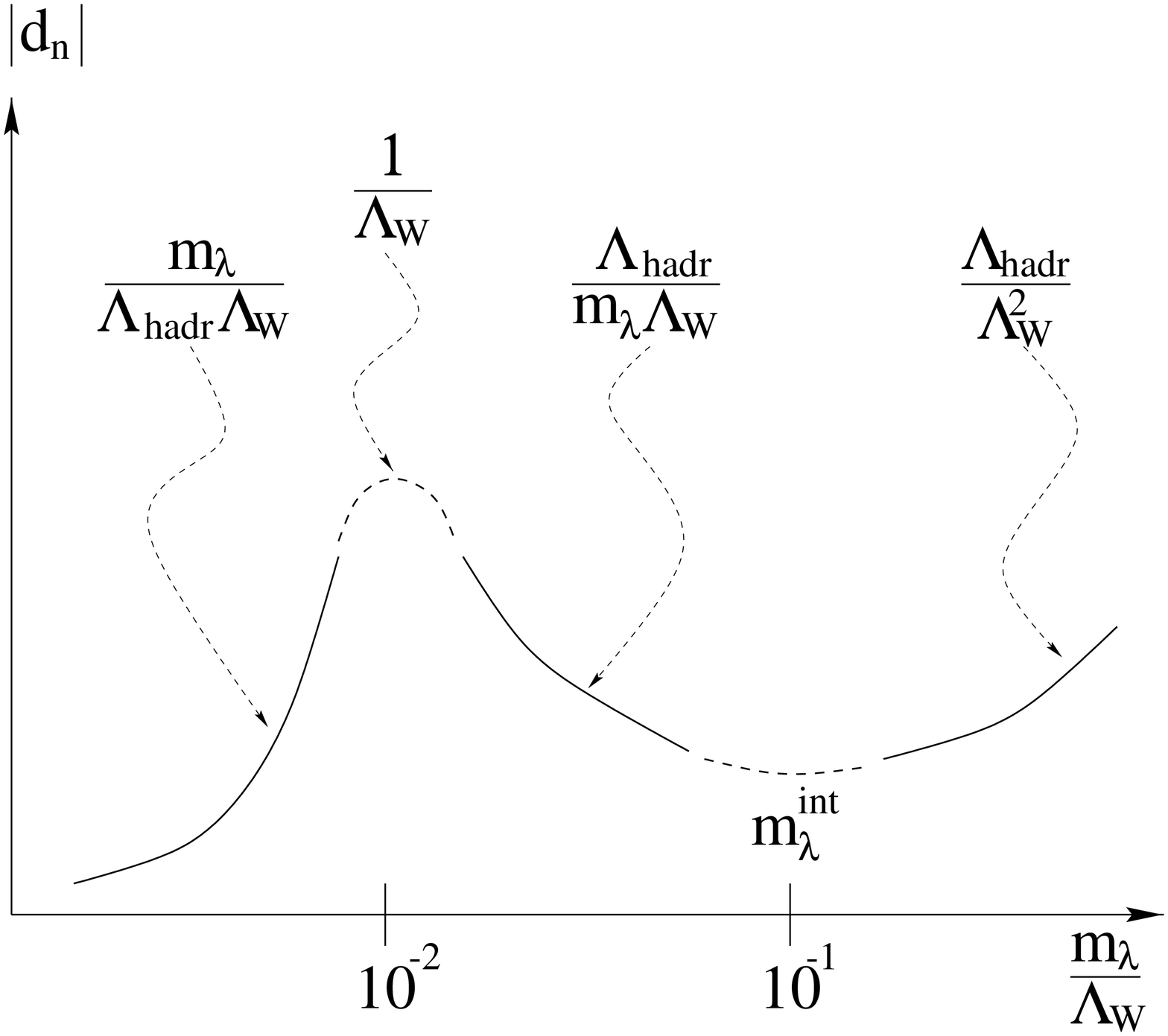, width=7cm, angle=0}
\end{center}  
\vskip .1in
\caption{\label{fig1} {\it Schematic behavior of the neutron EDM $d_n$ as
a function of the gluino mass. Lowering $m_\lambda $ from the SUSY
threshold there is an initial
suppression of $d_n$ due to the decrease of $d_i(m_\lambda)$ as
$m_\lambda$ decreases from $\La_W$ to the intermediate value
$m_\lambda^{\rm int}$. A further decrease of $m_\lambda$ in the  interval
from $m_\lambda^{\rm int}$ to $\Lambda_{ {\rm hadr}}$ leads to the
increase of  $d_n$ due to the contribution of the Weinberg operator,
induced by the gluino CEDM.  When $m_\lambda $ is
smaller than $\Lambda_{\rm hadr}$, $d_n$ receives a linear suppression
by $m_\lambda$.  }}
\end{figure}

\noindent
This enhancement
by the gluino CEDM (\ref{gcedm}) persists in the intermediate 
hierarchical regime $\Lambda_{\rm hadr} \ll m_\lambda \ll \Lambda_W$ where,
on integrating out the gluino, one generates a contribution 
to the Weinberg operator alluded to above that scales 
as  $1/(m_\lambda \Lambda_W)$. At a critical 
scale $m_\lambda=m_\lambda^{\rm int}<\Lambda_W$ these 
contributions will dominate over 
$d_i \sim \Lambda_{\rm hadr}m_\lambda/\Lambda_W^3$
and $d_n$ will start increasing while $m_\lambda$ decreases. 
As we will determine below, the scale
\be
 m_\lambda^{\rm int} \sim (6 - 12)~{\rm GeV}
\label{interm}
\ee 
sets an effective threshold for the maximal suppression of 
EDMs possible with this superpartner 
hierarchy\footnote{As an aside, we note that (perhaps 
surprisingly) a gluino mass of order  $m_\lambda^{\rm int}$ is still not
ruled out by direct constraints, and indeed has recently been revived
\cite{wagner}
in relation to the enhanced hadronic $b$-quark production observed at CDF and 
D$\slash \!\!\!0$ \cite{cdfd0}.}.

Our results suggest that at this scale the neutron EDM is 
still considerably larger than the experimental bound,
\be
 \De d_n(m_\lambda^{\rm int}) \sim (40 - 80)\,  d_n^{\rm exp},
\ee
unless the SUSY $CP$ phases are fine-tuned. Note that
both $m_{\lambda}^{\rm int}$ and  $d_n(m_\lambda^{\rm int})$ depend, in
addition, on possible inter-generational hierarchies for the squark masses. 
When the first generation of sfermions is taken to be heavier than $\Lambda_W$,
$m_{\lambda}^{\rm int}$ increases while $ d_n(m_\lambda^{\rm int})$ decreases.

Therefore, the Weinberg operator has an important role to play in 
minimizing the suppression possible within the light gluino regime. 
Note that for $m_\lambda \ll
\Lambda_{\rm hadr}$, the  $CP$-violating phase can be rotated to
$m_\lambda$ itself leading to a suppression of $d_n$ by
$m_\lambda/\Lambda_{\rm hadr}$ as one approaches the super Yang-Mills
limit. A schematic plot of the behaviour of $d_n(m_{\lambda})$ is
shown in Fig.~1.

In the next section we turn to the problem of estimating the 
contribution to $d_n$ induced by the Weinberg operator, justifying the
result (\ref{res}). Then, in Section~3 we describe in more detail the
calculation justifying the argument outlined above which 
uses the Weinberg operator to limit the
suppression of EDMs for light gluinos.

\section{Neutron EDM induced by the Weinberg operator}

Unlike the case of $d_n$ induced by the $\th$-term, or the EDMs and CEDMs
of quarks, where chiral loop  \cite{Crewther} and QCD sum rule--based
calculations \cite{PR} are available, the matrix element that relates
$d_n$ with the Weinberg operator is unknown. The standard 
estimate, first obtained by Weinberg \cite{wein1}, 
makes use of `naive dimensional analysis'
\cite{nda,nda2} which keeps track of dimensions, in terms of the generic
hadronic scale $\La_{\rm hadr}$, and Goldstone-mediated
interactions through the effective dimensionless coupling 
$\La_{\rm hadr}/f_\pi$. One finds \cite{wein1,wein2},
\be
\label{nda}
 d_n \sim e \frac{\Lambda_{\rm hadr}}{4
\pi}\ w  (\mu)\sim e~ 90~{\rm MeV}~w(\mu), \label{nda0}
\ee
at a low-energy normalization point $\mu$, 
taking $\Lambda_{\rm hadr} \sim 4\pi f_\pi\sim 1.2$ GeV. The
large value $\sim 4\pi$ for the coupling amounts to demanding that
loop corrections are qualitatively similar to the tree-level
terms at the matching scale. In the gluonic sector,
which is important here, this means that within the UV quark/gluon
description the relevant value of the gauge coupling is necessarily 
very large and consequently the inferred matching scale 
doesn't mesh easily with expectations from 
the chiral sector \cite{nda,nda2}. In the present context 
Weinberg \cite{wein1}, and many papers since \cite{wein2}, have, 
for the purpose of evaluating the gauge coupling, chosen a 
specific matching scale corresponding to $g_s = 4 \pi/\sqrt{6}$, 
or $\alpha_s\simeq 2$ (cf. $\alpha_s (1~{\rm GeV})\simeq 0.4$). 
If we adopt this normalization scale in (\ref{nda0}), and 
use (somewhat optimistically) the 1-loop anomalous dimension for 
$w$ \cite{evolve}, the relation  
$w(\mu(g_s=4\pi/\sqrt{6})) \simeq 0.4\, w(1 ~{\rm GeV})$
leads to the most commonly used estimate for $d_n(w)$: 
\be
 d^{(1)}_n \sim e\, 40 ~{\rm MeV}~ w(\mu=1 {\rm GeV}). 
\label{nda1}
\ee
We will avoid quoting a result for the dependence of $d_n$ on 
$w(\Lambda_{W})$, as there are additional threshold contributions
from $\tilde d_b$ and $\tilde d_c$ generated by top-stop-gluino loops,
which are in general model-dependent \cite{arnowitt}.

\begin{figure}
\begin{center}
\epsfig{file=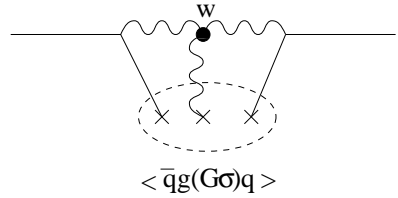, width=2.5cm, angle=270}
\end{center}  
\vskip .1in
\caption{\label{fig2} {\it Perturbative insertion of the Weinberg operator
into a quark line. The resulting correction  to the propagator is
proportional to $w\gamma_5\langle \bar q g_s(G\sigma)q\rangle$.  }}
\end{figure}

To get some intuition regarding the estimate (\ref{nda1}), we
can consider more carefully the loop-factors which
are effectively set to unity in (\ref{nda0}). For illustration, consider 
reducing the Weinberg operator to the EDM by `integrating out' the
gluons. This leads to an effective loop factor of $g_s^3/(4\pi)^4$ which 
reproduces (\ref{nda0}) provided we take $g_s \sim 4\pi$. 
One obtains a similar conclusion for the 
effective scale by considering the gauge kinetic term itself
\cite{nda}. As a consistent matching condition we might then 
choose $\mu=\mu(g_s=4\pi)$, leading to a result 
\be
 d^{(2)}_n  \sim e\,18 ~ {\rm MeV}~ w(\mu=1\, {\rm GeV}), \label{nda2}
\ee
which is half the size of (\ref{nda1}). Although both results 
(\ref{nda1}) and (\ref{nda2})
are consistent within the expected precision of NDA, it is clear
that independent quantitative calculations are needed to 
determine $d_n(w)$ to better than an order of magnitude.

As a quantitative test of the NDA estimates, we will now 
revisit the calculation of $d_n(w)$ using QCD sum rules, 
leading to a result that is a factor of 2 {\it smaller} than 
(\ref{nda1}) and consistent with (\ref{nda2}). To proceed,
we note first that the leading contribution to the EDM
from the operator product expansion (OPE) of the
nucleon current correlator in the presence of the 
source (\ref{wo}) exhibits a logarithmic infrared divergence.
This signals \cite{IS} the presence of additional operators,
required to resolve the divergence, whose contributions
are generally rather difficult to calculate directly. Therefore,
we will be content to regulate the log-divergent contributions
with an IR cutoff. These terms will then form the
basis of our estimates as they are correspondingly enhanced and 
thus provide the dominant contributions to the EDM.

We begin by noting that the Weinberg operator allows for 
a perturbative insertion into the quark propagator. The leading
$CP$-odd correction is described by the diagram shown in Fig.~2, and
standard manipulations \cite{nsvz} lead to the following result:
\be 
iS(p)= \fr{i \psl}{p^2} + \fr{i g_s w}{8
p^4} \ga_5 \langle \bar q g_s(G\sigma) q\rangle,
\label{sp}
\ee 
where the value of the quark-gluon condensate is given by \cite{bi}
\be
\langle \bar q g_s (G\sigma) q\rangle = m_0^2 \vert\langle \bar q
q\rangle\vert \simeq 0.8 {\rm GeV}^2  \langle \bar q  q\rangle, 
\ee
with $\langle \bar qq \rangle = - (230\ {\rm MeV})^3$. It is
the $1/p^4$ momentum dependence in the second term of Eq.~(\ref{sp})
which leads to the logarithmic infrared divergence alluded to above
in the correlator of two nucleon currents.
This signals the breakdown of the OPE, but also singles out this
insertion as providing the dominant effect which we will use
in calculating $d_n(w)$. The ambiguity of the infrared log does
of course render the result less reliable than the corresponding 
determination of $d_n(d_i,\widetilde{d}_i)$ \cite{PR}, but 
nonetheless sufficient for our estimates.

The insertion present in the second term in (\ref{sp}) behaves as 
a ``soft $\gamma_5$-mass''. Indeed, while irrelevant for large $p^2$,
at hadronic scale momenta it mimics the existence of an effective $CP$-odd
mass of order 
\be 
m_5^{\rm eff} \simeq \fr{g_s w}{8 \Lambda_{\rm hadr}^2}
\langle \bar q g_s(G\sigma) q\rangle \sim ( 120~{\rm MeV}-160~{\rm MeV})^3\, w
\label{m5}
\ee 
where $\Lambda_{\rm hadr}$ is the effective hadronic scale, which we take
to lie in a range from $m_{\rh}$ up to $4\pi f_{\pi}$, with previous
results \cite{PR} suggesting that the lower end of this range is most relevant
for EDM observables. This determines the neutron
EDM according to the scaling relation \cite{PR},
\be 
 d_n \sim e\fr{|m_5^{\rm eff}|}{\La_{\rm hadr}^2} \sim e~(1.5-7)~{\rm MeV}~
w(\Lambda_{\rm hadr}),
\label{est1} 
\ee
where the large range in this estimate arises from the allowed
variation in $\La_{\rm hadr}$. 

This result is 5--10 times smaller than the conventional NDA estimate
(\ref{nda1}). This is actually not too surprising once we recall that
a priori $d_n(w)$ should be of $O(\qq^0)$ in the chiral limit, while 
the contribution in (\ref{est1}) is $O(\qq)$ and thus may indeed be 
subleading. To test this one can
consider an explicit sum-rules based estimate \cite{PRinprep} utilizing the 
insertion (\ref{sp}). One finds that for the natural chirally
invariant Lorentz structure, $\{\psl, (F\si)\gamma_5\}$ \cite{PR}, 
the tractable contributions are of $O(\qq^2)$ and render a result for 
$d_n$ within the range (\ref{est1}). Previous experience \cite{PR} would 
suggest that the terms of $O(\qq^2)$ are sub-dominant, but
unfortunately the (a priori) leading contributions
of $O(\qq^0)$ for $d_n(w)$ are intractable in this direct 
approach due to the presence of unknown condensates. 

This analysis suggests that the range (\ref{est1}) 
might represent an underestimate of $d_n(w)$. A natural path to
follow is to consider the sum-rules in chirally-variant channels
such as $(F\si)$ or $\psl(F\si)\psl$ from which one can still
extract $d_n(w)$ along the lines considered previously for 
$d_n(\theta)$ \cite{Henley}. A convenient means of estimating 
$d_n(w)$ in this vein is to calculate the $\ga_5$--rotation of the
nucleon wavefunction induced by the Weinberg operator and 
determine $d_n$ in terms of the corresponding rotation of the 
neutron anomalous magnetic moment $\mu_n$:
\begin{eqnarray} 
d_n\sim \mu_n\,\fr{\langle N\vert
\fr{w}{3}(GG\tilde G)\vert N\rangle} {m_n\bar N i\gamma_5 N}.
\label{BU}
\end{eqnarray}
This approach was considered previously by Bigi and Uraltsev \cite{BU} who 
estimated $\langle N\vert (GG\tilde G)\vert N\rangle$ in terms of
$\langle N\vert GG \vert N\rangle$ and the corresponding vacuum
condensates. 

We can follow this route and perform a more explicit calculation
by evaluating the `$\gamma_5$'  term in the standard
mass sum-rule correlator of the two nucleon currents. 
For the conventional choice of the 
Ioffe interpolating current for the neutron $\et$ \cite{sumrules}, we obtain
at leading order,
\bea 
  && \int d^4x e^{ip\cdot x}\langle \eta(0) \bar
 \eta(x)\rangle  = \frac{1}{16\pi^2} 
 p^2 \ln (-\La_{\rm UV}^2/p^2) \langle \bar{q} q \rangle  
 \;\;\;\;\; \nonumber\\
  && \;\;\;\;\;\;\;\;\;\;\;\; \times 
 \left [ 1 + i \gamma_5 \fr{3 g_s w}{32
 \pi^2} m_0^2 \ln{(-p^2/\mu_{IR}^2)} \right ] + \cdots .
\label{2point}
\eea 
It is the relative coefficient between the structures 
{\bf 1} and $i \gamma_5$ that determines the chiral rotation
and consequently enters into the estimate of \cite{BU}.
From (\ref{BU}) and (\ref{2point}) we obtain
\be
  d_n \simeq \mu_n  \fr{3 g_s m_0^2}{32 \pi^2} w\ln{(M^2/\mu_{IR}^2)} 
 \simeq e~ 22~{\rm MeV}~ w(1~{\rm GeV}),
\label{est2}
\ee 
where we took $M/\mu_{IR} = 2$ and $g_s = 2.1$. It is important to note that 
the estimate (\ref{est2}) arises at $O(\qq^0)$, which we
would expect to be dominant, and is indeed 
considerably larger than the estimate (\ref{est1}). A more involved 
calculation of the nucleon current correlator in an external 
electromagnetic field \cite{PRinprep} reveals additional 
contributions to $d_n(w)$, but the overall result remains quite 
close to (\ref{est2}). Additional induced corrections, from Peccei-Quinn
relaxation, would also be subleading \cite{BU} as they cannot
contribute at $O(\qq^0)$.

The only other QCD sum-rules estimate of $d_n(w)$ that we are 
aware of was made by Khatsimovsky \cite{Kha} who considered a high
order term in the OPE proportional to the dimension-eight operator
$F(GG\tilde G)$. An estimate of the nonlocal correlator, 
$\int d^4 x \langle 0\vert T\{(GG\tilde G)(0),(GG\tilde G)(x)\} \vert
0\rangle $ produced a result for $d_n(w)$ similar to (\ref{nda0}).
However, combinatoric factors were ignored in this calculation
which clearly reduce the result to a value consistent with -- or
somewhat smaller than -- (\ref{est1},\ref{est2}). In practice 
a precise calculation along these lines does not appear feasible, as 
multiple perturbative insertions of the gluon field strength into a 
quark line generally leads to power-like infrared divergences \cite{IS},
signifying the breakdown of the OPE. 

Putting these results together, and ignoring the lower range
of (\ref{est1}) for the reasons discusssed above, we find the preferred
range for $d_n(w)$,
\begin{eqnarray}
 d_n(w) \sim e\,(10-30)~{\rm MeV}\, w(1~{\rm GeV}),
\end{eqnarray}
which is a factor of two smaller than the conventional NDA estimate
(\ref{nda1}), and consistent with (\ref{nda2}). This
result will be discussed in more detail elsewhere \cite{PRinprep}, but
we turn now to a regime of the SUSY parameter space for which this 
contribution to $d_n$ is nonetheless very significant.

\section{Enhancement via gluino color EDM}

As described and schematically illustrated in Section~I, the neutron EDM is
particularly enhanced in the domain $\Lambda_{\rm hadr}\leq m_\lambda
\ll \Lambda_{W}$ where the gluino develops a color EDM via a
top--stop loop \cite{apostolos},
\begin{eqnarray}
\widetilde{d}_{\lambda}(\Lambda_W) &=& - \frac{g_s^3(\Lambda_W)}{32 \pi^2}\
\frac{m_t}{M_{\widetilde{t}_1}^{2}}\ \sin (2 \theta_{\widetilde{t}})\
\sin \delta_{t}\ \nonumber\\
 && \;\;\;\;\;\; \times \left[f_g\left(
\frac{m_t^2}{M_{\widetilde{t}_1}^{2}}\right)-\frac{M_{\widetilde{t}_1}^{2}}
{M_{\widetilde{t}_2}^{2}}
f_g\left(\frac{m_t^2}{M_{\widetilde{t}_2}^{2}}\right)\right], \label{tsloop}
\end{eqnarray}
with $\delta_t=\mbox{Arg}[A_t-\mu^{\star} \cot\beta]$ and
$f_g(y)=(1-y+\ln(y))/(1-y)^2$.  Note that this expression is
independent of $m_\lambda$ and scales as $1/\Lambda_W$.  The
corresponding contribution to the Weinberg operator 
\cite{wein1,wein2,dai,bquark},
\begin{eqnarray}
\label{shift}
\Delta w (m_\lambda)= - \frac{3 g_s^2(m_\lambda)}{32\pi^{2}}\
\frac{\widetilde{d}_{\lambda}(m_\lambda)}{m_\lambda}
\end{eqnarray}  
scales as $1/m_\lambda\Lambda_W$. It is worth noting that in addition
to the obvious enhancement by a factor of $\Lambda_W/m_\lambda$
relative to the standard scenario \cite{dai}, the gluino CEDM--induced
shift of the Weinberg operator is also enhanced relative to that
induced by $c$ or $b$ quarks which is of order $1/\Lambda_W^{2}$
\cite{bquark,arnowitt}. 

The normalization of $\Delta w$ at the 
hadronic scale involves running the gluino CEDM 
from $\Lambda_W$ down to the gluino mass threshold, and 
subsequent running of $w$ down to $\Lambda_{\rm hadr}$. For
completeness, we give the 1-loop $\beta$ function coefficient,
$\beta_0=11-2 n_{\lambda}- 2 n_q/3$, where $n_x$ stands for the 
number of light $x$ particles at the scale under
concern. Besides this, the anomalous 
dimensions of $\widetilde{d}_{\lambda}$
and $w$ are given, respectively, by 
$\gamma^{\lambda}=-18+\beta_0$ and
$\gamma^{W}=-36+3 \beta_0$. The latter has been 
computed in \cite{evolve}, and the
computation of the former is similar to that of the
quark color EDMs \cite{eski}.

We now illustrate numerically the impact of light gluinos on the
neutron EDM using the range for $d_n(w)$ in (\ref{res}). 
Using (\ref{tsloop}) and (\ref{shift}), we can write 
\be
\De d_n \sim  100\  \sin\de_t\ \frac{{\rm (4 - 12)\, GeV}}{m_{\lambda}}
 \ d_n^{\rm exp}, 
\ee
where we have taken $M_{\widetilde{t}_1}=200\ {\rm GeV}$, 
$M_{\widetilde{t}_2}=700{\rm GeV}$, $\theta_{\widetilde{t}}=\pi/4$, 
and the current experimental bound on the neutron EDM is 
$d_{n}^{\rm exp}< 6 \times 10^{-26}\ {\rm e.cm}$ \cite{exp}. 
The final results are presented in Fig.~3, where we have chosen the mid-value
$d_n = 20\,$MeV$\,w$ in (\ref{res}). The solid curve stands for 
$m_{\lambda}= 1\ {\rm GeV}$ (with $d_n/d_n^{\rm exp}\approx 980$ 
at $\delta_t=\pi/2$), the dashed curve for $m_{\lambda}= m_b$ 
(with $d_n/d_n^{\rm exp}\approx 170$ at $\delta_t=\pi/2$), and 
the dot--dashed curve for $m_{\lambda}= 20\ {\rm GeV}$ 
(with $d_n/d_n^{\rm exp}\approx 30$ at $\delta_t=\pi/2$). 
Thus, for light gluinos, where $m_{\lambda}\sim (1 - 4)\,{\rm GeV}$,
one finds that $d_n(w)$ exceeds the experimental bound by at least
two orders of magnitude throughout the entire preferred range in (\ref{res})
unless the SUSY phases are tuned such that $\delta_t\simlt 10^{-2}$.

\begin{figure}[t]
\begin{center}
\epsfig{file=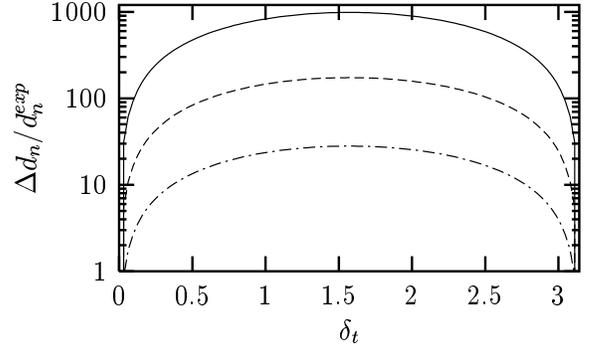, width=8cm, angle=0}
\end{center}  
%\vskip .1in
\caption{\label{fig3} {\it The $\delta_t$ dependence of the gluino contribution
to $d_{n}$ for $m_{\lambda}=1\ {\rm GeV}$ (solid), 
$m_{\lambda}=m_b$ (dashed),
and $m_{\lambda}=20\ {\rm GeV}$ (dot--dashed).}}
\end{figure}

Of particular interest is the maximal suppression that one can achieve
for the EDM in this hierarchical regime with light gluinos. We denote
by $m_{\lambda}^{\rm int}$ the critical scale at which the 1-loop 
contribution induced by quark EDMs (and CEDMs) is approximately equal to 
the contribution associated with the Weinberg operator discussed here.
Choosing the soft-breaking parameters in the first generation
of squarks to be ${\cal{O}}(200\ {\rm GeV})$, and assuming no accidental
cancellations, we find 
\be
 m_{\lambda}^{\rm int} \sim (6-12)\, {\rm GeV} 
\ee
accounting for the range in (\ref{res}), 
for which the (minimal) correction to the EDM is 
approximately, 
\be
 \De d_n(m_\lambda^{\rm int}) \sim (40-80)\, d_n^{\rm exp},
\ee
which still exceeds the experimental bound by at least
an order of magnitude unless the $CP$-odd phases are small.

It is interesting to compare our estimates for $d_n$ with those 
one obtains when the gluino is heavy, $m_{\lambda}\sim \Lambda_W$.  
In this case, the Weinberg operator is first generated at the weak
scale at two-loop order \cite{dai}. On including the contributions arising at
the $b$--quark and $c$--quark thresholds, one finds that $d_n$
obtained via (\ref{nda}) only exceeds $d_n^{\rm exp}$ by at most one 
order of magnitude \cite{arnowitt}. Consequently, the light gluino scenario
actually induces a larger contribution to $d_n$ via the 
color EDM of the gluino. Thus, while it is possible to suppress the
one--loop contributions to the EDMs of leptons and hadrons by taking 
light gauginos \cite{Austrians}, the induced contribution to the
Weinberg operator means that the constraints on the SUSY $CP$-odd phases
are not correspondingly relaxed.

We thank Louis Clavelli, Glennys Farrar and Oleg Lebedev 
for useful discussions. The work of D.D. was supported in part 
by DOE grant DE-FG02-94ER40823.

\end{document}